\documentclass[10pt,twocolumn,letterpaper]{article}

\usepackage{cvpr} 
\usepackage{times}
\usepackage{epsfig}
\usepackage{graphicx}
\usepackage{amsmath}
\usepackage{amssymb}
\usepackage{soul}
\usepackage{mathtools}
\usepackage{multirow}
\usepackage{booktabs}
\usepackage{array}
\usepackage{subcaption}
\usepackage{hyperref}
\usepackage{stfloats}

\usepackage{footnote}
\makesavenoteenv{table}
\makesavenoteenv{figure}

\newcommand*{\X}{\mathcal{X}}

\newcommand*{\Z}{\mathcal{Z}}

\begin{document}

\title{WaFusion: A Wavelet-Enhanced Diffusion Framework for\\ Face Morph Generation}

\author{Seyed Rasoul Hosseini, Omid Ahmadieh, Jeremy Dawson and Nasser Nasrabadi\\
West Virginia University\\
Morgantown, USA\\
{\tt\small sh00111, oa00023@mix.wvu.edu, jeremy.dawson@mail.wvu.edu, nnasrab1@gmail.com}
}

\maketitle
\thispagestyle{empty}

\begin{abstract}
   Biometric face morphing poses a critical challenge to identity verification systems, undermining their security and robustness. To address this issue, we propose WaFusion, a novel framework combining wavelet decomposition and diffusion models to generate high-quality, realistic morphed face images efficiently. WaFusion leverages the structural details captured by wavelet transforms and the generative capabilities of diffusion models, producing face morphs with minimal artifacts. Experiments conducted on FERET, FRGC, FRLL, and WVU Twin datasets demonstrate WaFusion's superiority over state-of-the-art methods, producing high-resolution morphs with fewer artifacts. Our framework excels across key biometric metrics, including the Attack Presentation Classification Error Rate (APCER), Bona Fide Presentation Classification Error Rate (BPCER), and Equal Error Rate (EER). This work sets a new benchmark in biometric morph generation, offering a cutting-edge and efficient solution to enhance biometric security systems.  
\end{abstract}

\section{Introduction}

Owning to non-intrusive nature, ease of use, and broad acceptance \cite{icao20159303}, Facial Recognition Systems (FRS) are widely adopted in security-related applications, particularly as the primary biometric for electronic Machine-Readable Travel Documents (eMRTD) \cite{makrushin2020simulation, wolberg1998image, zope2017survey}. However, the increased reliance on FRS exposes them to face morphing attacks, \ie meticulously blending the facial features of two individuals to create an image resembling both identities \cite{godage2022analyzing, scherhag2017biometric}. These morphed images undermine FRS by increasing the False Acceptance Rates (FAR), deceiving human verifiers, and risking unauthorized access by malicious actors to secure facilities \cite{Ferrara2014TheMP, robertson2017fraudulent}. As morphing tools become more accessible, even individuals with limited technical skills can generate high-quality morphs, exacerbating these risks \cite{bowyer2004face}. 

While efforts to counter face morphing often focus on detection methods integrated into biometric pipelines that identify morphed images \cite{Venkatesh2020FaceMA}, these approaches struggle against high-quality morphs, notably, those generated by advanced deep learning techniques \cite{ramachandra2024multispectral, zhang2024generalized}. This highlights the urgent need for improved morph generation techniques to evaluate and intensify detection algorithms. 

Morphing methods typically fall into two categories: 1) landmark-based approaches \cite{ferrara2019decoupling} and 2) deep learning-based techniques \cite{damer2018morgan}. Landmark-based methods rely on a structured pipeline involving landmark detection, warping, and blending \cite{cootes2001active, opencv, facemorpher}. While effective, these methods often require manual adjustments and are prone to obvious visual inconsistencies. Conversely, deep learning-based approaches, such as Generative Adversarial Networks (GANs), generate morphs in an end-to-end manner \cite{karras2019style}. GANs have become a prominent choice due to their ability to produce high-quality images through adversarial training \cite{goodfellow2020generative}. 
However, GAN-based methods face challenges such as mode collapse, geometric distortion, and sensitivity to dataset quality, often leading to artifacts or vulnerability to adversarial-style variations \cite{ashwini2023generation, saadabadi2024boosting, xiao2021tackling}. 

Recently, Diffusion Probabilistic Models (DPMs) have emerged as a robust alternative to GANs for image synthesis, offering enhanced diversity and fidelity \cite{ho2020denoising}. These models iteratively refine noisy data to model complex distributions, enabling the generation of high-quality images \cite{hamza2022comprehensive, kim2024diffusion}. The Denoising Diffusion Probabilistic Model (DDPM), introduced by Ho et al. \cite{ho2020denoising}, achieves remarkable image quality through iterative noise reduction. Building on this, the Denoising Diffusion Implicit Model (DDIM) improves efficiency by reducing the number of sampling steps while maintaining high synthesis quality \cite{song2020denoising}. Nevertheless, diffusion models face notable challenges, requiring substantial computational resources and extended runtimes. They also struggle with fine-grained control over morphological traits, making them less practical for specific tasks \cite{dhariwal2021diffusion, rombach2022high, xiao2021tackling}. These limitations underscore the need for approaches combining diffusion models' strengths while addressing their challenges, such as improving computational efficiency and feature control.

This work aims to generate high-quality face morphs while preserving critical image details efficiently. Despite advancements in state-of-the-art (SOTA), including GANs and diffusion models, existing methods face significant limitations. To overcome these challenges, we need a hybrid approach that integrates complementary techniques. 

To this end, we propose a novel hybrid framework that combines Discrete Wavelet Transform (DWT) and diffusion models. Wavelet decomposition excels in preserving essential image textures by analyzing images at multiple frequency bands \cite{daubechies1992ten, mallat1989theory}, while diffusion models ensure high-fidelity synthesis \cite{ho2020denoising, song2020denoising}. In our approach, 
the Low-low (LL) sub-band, which captures critical image features, is extracted through wavelet decomposition and processed using diffusion models to generate realistic morphed components. The Inverse Wavelet Transform (IWT) then combines the morphed LL sub-band with the high-frequency sub-bands from the original image, producing the final morph. By using the complementary strengths of these techniques, our method effectively addresses the limitations of existing approaches, enabling the generation of high-quality morphs with enhanced efficiency.  

The primary contributions of this study are as follows: 
\begin{itemize}
    \item An innovative approach using wavelet decomposition for biometric morphing, enhancing image quality while preserving essential features. 
    \item The ability to generate high-resolution morphs without increasing computational costs compared to conventional baseline methods. 
    \item A comprehensive evaluation of the proposed framework against SOTA techniques across diverse datasets, demonstrating its superior quality, efficiency, and robustness. 
\end{itemize}

\section{Related Works}

The risks posed by face morphing attacks are highlighted by Ferrara \textit{et al.} \cite{Ferrara2014TheMP}, who manually created morphed images using GIMP, an open-source image editor. These images could deceive Automated Border Control (ABC) systems despite showing minimal artifacts, but their manual nature lacked scalability for large-scale dataset production. Landmark-based algorithms such as Facemorpher \cite{facemorpher} and OpenCV \cite{opencv} subsequently automated morph generation, using warping and splicing to produce morphs efficiently. However, these methods often introduce artifacts in high-frequency regions like the iris and facial contours \cite{zhang2021mipgan}, limiting their effectiveness for realistic morph generation. 

Recent developments in GANs have significantly improved morph quality by automating synthesis and minimizing perceptual inconsistencies \cite{damer2021regenmorph, GomezBarrero2022OnTD, sarkar2022gan}. Damer \textit{et al.} \cite{damer2018morgan} introduced MorGAN, one of the first GAN-based morphing methods, which blends two identities into a single morphed identity. Subsequent architectures such as StyleGAN2 \cite{sarkar2020vulnerability} and MIPGAN-II \cite{zhang2021mipgan} further enhanced morph quality and reduced artifacts. These advancements have driven further research in the field of face morph generation \cite{price2022landmark, venkatesh2020can}. However, GAN-based approaches face persistent challenges, such as mode collapse and dataset sensitivity, often resulting in artifacts \cite{roich2022pivotal}.  

DPMs have emerged as a promising alternative to GANs, offering superior diversity and fidelity \cite{sohl2015deep}. 
Ho \textit{et al.} \cite{ho2020denoising} demonstrated the potential of DPMs for generating highly detailed images through iterative denoising and effectively addressing mode collapse. 
Diffusion autoencoders extended these capabilities by disentangling semantic and random data, allowing for more controlled image generation \cite{Preechakul2021DiffusionAT}. 
Building on this, recent methods have applied diffusion models specifically to face morphing.
DiffMorpher \cite{zhang2024diffmorpher} enables full-image semantic interpolation but incurs high computational costs. 
Fast-DiM \cite{blasingame2024fast} improves efficiency by reducing the number of sampling steps but does not enhance morph quality or structure preservation. 
Blasingame \textit{et al.} \cite{blasingame2024leveraging} proposed a strong morphing attack pipeline based on diffusion, yet it lacks control over fine-grained facial features.
Despite these advances, computational demands and limited structural fidelity remain open challenges.
In contrast, WaFusion introduces a hybrid wavelet-diffusion approach that selectively processes only low-frequency components, achieving efficient morphs while maintaining critical facial structure.

Wavelet-based methods have proven effective in enhancing both the quality and robustness of face morphs. O'Haire \textit{et al.} \cite{OHaire2021AdversariallyPW} employed wavelet transformations to generate morphed faces, while Huang \textit{et al.} \cite{huang2023wavedm} applied them for image restoration in morphing contexts. More recently, Hybrid approaches such as Phung \textit{et al.} \cite{phung2023wavelet} have combined wavelet decomposition with diffusion models, leveraging texture analysis for high-fidelity synthesis with reduced computational cost. These advancements demonstrate the adaptability of wavelet-based frameworks in addressing prominent challenges in morph generation. 

As face morphing techniques evolve, Morphing Attack Detection (MAD) has become a critical research area. Early works \cite{Raghavendra2016DetectingMF, scherhag2019detection} emphasized the challenge of distinguishing morphs from bona fide images. More recent approaches leverage diffusion models to detect morphs as out-of-distribution samples \cite{ivanovska2023face}, deep face embeddings from models like MagFace \cite{kessler2024towards}, and multispectral imaging for enhanced detection \cite{ramachandra2024multispectral}. These works demonstrate how progress in morph generation drives advances, reflecting the co-evolution of attack and defense strategies.

While landmark-based methods offer simplicity and control, they often fail to capture complex facial structures and expressions. GAN- and diffusion-based techniques significantly improve realism and diversity, yet they remain susceptible to artifacts and high computational cost \cite{Venkatesh2020FaceMA}. Recent hybrid approaches that integrate wavelet decomposition with generative models show potential in addressing these limitations. Building on this direction, our approach combines wavelet decomposition with diffusion autoencoders to selectively enhance structural fidelity while reducing generation overhead.

\section{Methodology}

\begin{figure*}[t]
\begin{center}
    \includegraphics[width=0.9\linewidth]{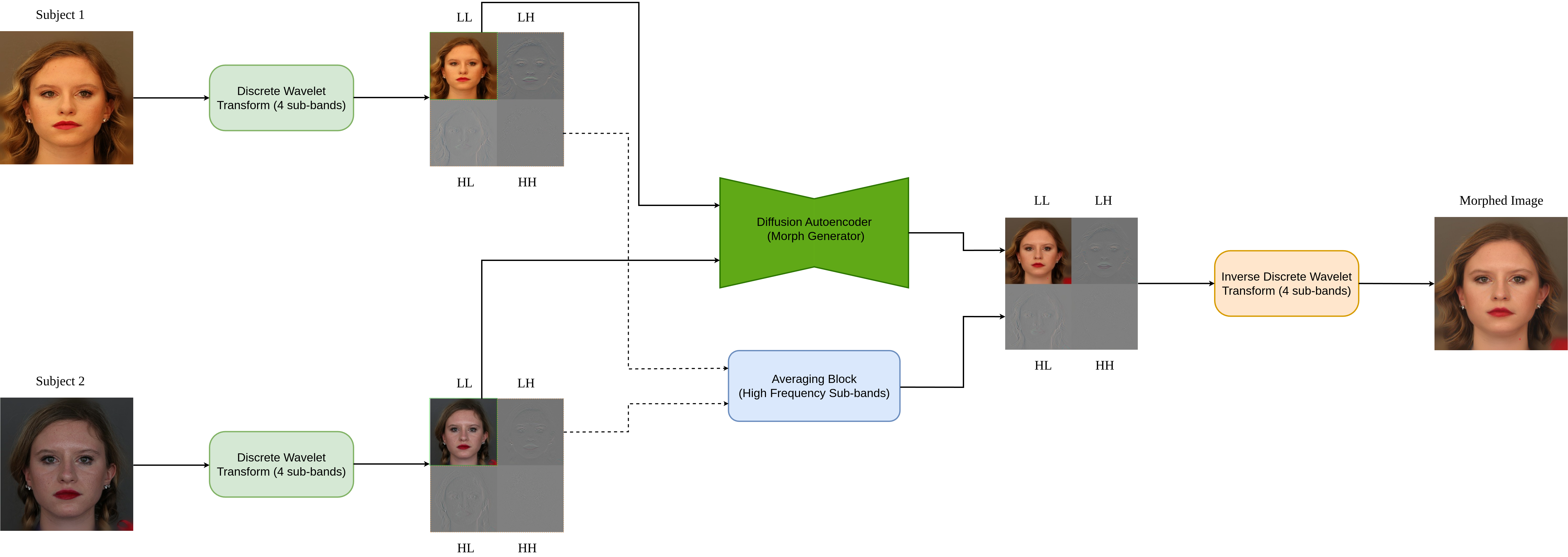}
\end{center}
   \caption{Overview of the WaFusion framework for face morphing. The input images, Subject 1 and Subject 2, are first aligned and decomposed into four sub-bands (LL, LH, HL, HH) using wavelet decomposition. The LL sub-bands, highlighted in the green dotted box, are processed through the diffusion model to generate the morphed LL sub-band. The remaining sub-bands, shown by orange dots, are averaged. The final morph is reconstructed using the inverse wavelet transform, combining both morphed and averaged components.}
\label{fig:WaFusion}
\end{figure*}
\subsection{Wavelet Transform}

The wavelet transform is a foundational component of our model, enabling efficient decomposition of images into their low-frequency approximations and high-frequency details \cite{mallat1989theory}. This hierarchical process captures coarse structure in the Low-Low (LL) sub-band, while encoding finer details , \emph{e.g.}, vertical, horizontal, and diagonal edges in the remaining sub-bands, namely Low-High (LH), High-Low (HL), and High-High (HH). Wavelet decomposition preserves essential structural and textural information while reducing the spatial dimensions, facilitating efficient downstream processing. 

We use the Haar wavelet for its simplicity and computational efficiency. The decomposition employs low-pass (L) and high-pass (H) filters, resulting in four sub-bands: \textbf{$X_{ll}$} represents a coarse, low-frequency approximation for the image, capturing its overall structure, \textbf{$X_{lh}$} encodes high-frequency details in the horizontal direction, highlighting fine horizontal edges, \textbf{$X_{hl}$} encodes high-frequency details in the vertical direction, focusing on vertical edges, and \textbf{$X_{hh}$} encodes high-frequency details in the diagonal direction, representing texture and diagonal edges. 


These sub-bands are represented in Fig. \ref{fig:WaFusion}, where the LL sub-band is highlighted as the primary source of structural information for morph generation, while the high-frequency sub-bands contribute fine-grained textures.
The Haar wavelet transform of an image $X \in \mathbb{R}^{H\times W}$ is computed as:

\begin{equation} \label{eq2}
X_{pq} = F_p^\top \cdot X \cdot F_q,  \hspace{5mm}   p, q \in \{L, H\},
\end{equation}
yielding four $\frac{H}{2} \times \frac{W}{2}$ sub-bands.
To reconstruct the original image, the IWT recombines these sub-bands, ensuring no loss of fidelity \cite{mallat1989theory}. 

In our framework, the LL sub-band is processed using a diffusion model to generate the morphed component. The high-frequency sub-bands (LH, HL, and HH) are averaged and fused with the morphed LL sub-band using the IWT to produce the final high-resolution morph. The wavelet analysis was implemented using the PyWavelets package \cite{pywt}, ensuring efficient computation and detailed texture preservation. 

\begin{figure*}[t]
\begin{center}
    \includegraphics[width=0.9\linewidth]{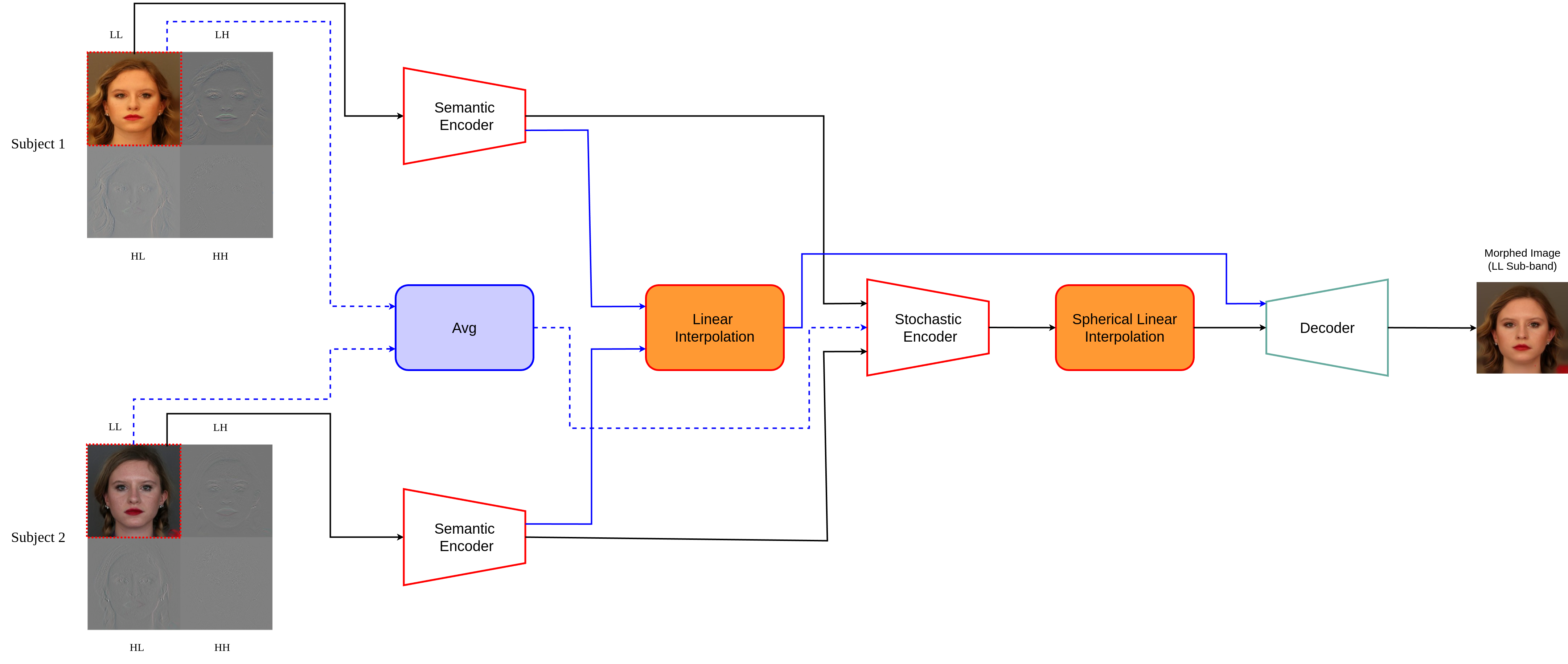}
\end{center}
   \caption{Architecture for diffusion-based morph generation using Diffusion Autoencoders. LL sub-bands (red dashed lines) from Subject 1 and Subject 2 are processed through semantic encoders and an averaging block. Outputs are passed through stochastic encoders, followed by interpolation. Semantic features undergo linear interpolation, while stochastic features use spherical linear interpolation. The combined features are decoded to generate the morphed LL sub-band.\protect\footnotemark[1]}
\label{fig:diffdiagram}
\end{figure*}
\footnotetext[1]{Note: The subjects shown in Figure 2 are identical twins from the WVU Twin dataset, which explains the high visual similarity between the bona fide images. This characteristic applies to all image pairs generated from the WVU Twin dataset.}
\subsection{Diffusion Model}
Diffusion-based generative models progressively transform a noise map, initialized from a standard Gaussian distribution $\mathcal{N}(\mathbf{0}, \mathbf{I})$, into a clean image through a series of denoising iterations. First introduced by Ho \textit{et al.} \cite{ho2020denoising}, these models rely on a learned denoising function, $\epsilon_\theta(x_{t}, t)$, which predicts the noise added to an image at step $t$. By iteratively removing this noise, the models reconstruct the original image $x_{0}$, starting from a fully noisy sample $x_{t}$. This denoising function is implemented using a UNet architecture \cite{ronneberger2015u}, enabling high-quality image generation. Moreover, the approach simplifies the variational lower bound on the data's marginal log-likelihood, a technique that has gained significant adoption in the field \cite{song2020denoising, nichol2021improved}.

In the forward process, Gaussian noise is added to $x_0$ over $T$ steps, modeled as:

\begin{equation}
    q(x_t \mid x_{t-1}) = \mathcal{N}((\sqrt{1 - \beta_t})x_{t-1}, \beta_t \mathbf{I}),
\end{equation}
where $\beta_t$ controls the noise schedule. The noisy image $x_t$ at step $t$ can also be expressed directly in terms of $x_0$:

\begin{equation}
    q(x_t \mid x_0) = \mathcal{N}(\sqrt{\alpha_t} x_0, (1-\alpha_t)\mathbf{I}),
\end{equation}
with $\alpha_t = \prod_{s=1}^t (1-\beta_s)$. The reverse process aims to remove this noise, reconstruct $x_{0}$, modeled as:

\begin{equation}
    p(x_{t-1} \mid x_t) = \mathcal{N}(\mu_\theta(x_t, t), \sigma_t),
\end{equation}
where $\mu_\theta$ is a learned mean function and $\sigma_t$ is predefined \cite{ho2020denoising}. This reverse process enables the stepwise reconstruction of $x_0$ from $x_T$.

Building on this foundation, Song \textit{et al.} \cite{song2020denoising} introduced the DDIM, which modifies the reverse process to be deterministic. DDIM generates samples using the following update rule:

\begin{equation}
\begin{aligned}
x_{t-1} = \sqrt{\alpha_{t-1}} \left( \frac{x_{t} - \sqrt{1 - \alpha_t} \epsilon_\theta^{t}(x_{t})}{\sqrt{\alpha_t}} \right) \\ & \hspace{-1cm} + \sqrt{1 - \alpha_{t-1}} \epsilon_\theta^{t}(x_{t}),
\label{eq:gen}
\end{aligned}
\end{equation}
allowing faster sampling and more precise control over the generative process without altering the marginal distribution. DDIM further introduces an inference distribution:

\begin{equation}
\begin{aligned}
q(x_{t-1} \mid x_t, x_0) = \\
& \hspace{-2cm} \, \mathcal{N}\Bigg( \sqrt{\alpha_{t-1}} x_0 + \sqrt{1 - \alpha_{t-1}} 
& \cdot \frac{x_t - \sqrt{\alpha_t} x_0}{\sqrt{1 - \alpha_t}}, \mathbf{0} \Bigg),
\label{eq:ddimq}
\end{aligned}
\end{equation}
which retains the core principles of DDPM while enabling more efficient and deterministic sample generation.

Despite their advantages, DPMs have limitations. The latent variable $x_{T}$, representing the starting point of the reverse process, lacks high-level semantic information, making it less interpretable for applications requiring meaningful feature manipulation, such as face morphing. Preechakul \textit{et al.} \cite{Preechakul2021DiffusionAT} explored this limitation and proposed methods to disentangle semantic and random data, highlighting the importance of enriching the latent variable's representation for downstream applications.

Our hybrid approach builds on the foundational strengths of DPMs while addressing their inherent limitations. By integrating structural precision with detailed feature control, this methodology enables the generation of high-quality, realistic face morphs, while enhancing both interpretability and computational efficiency.

\subsection{Proposed Method}

The proposed WaFusion framework for face morph generation is depicted in Fig. \ref{fig:WaFusion}, which provides an overview of its key components and workflow. The process begins with the decomposition of two bona fide input images, \ie distinct identities, into four wavelet sub-bands: LL, LH, HL, and HH. This decomposition, performed using a single-level Haar transform, which separates low-frequency approximations (LL) from high-frequency details (LH, HL, HH).
The LL sub-bands, which capture the structural essence of the images, are then processed by the morph generation block, implemented using diffusion autoencoders \cite{Preechakul2021DiffusionAT}, to generate the morphed LL sub-band. Finally, the inverse wavelet transform fuses the morphed LL sub-band with the averaged high-frequency sub-bands from subject 1 and subject 2, reconstructing the final morphed image with high fidelity and detail.

Our diffusion autoencoders, as illustrated in Fig. \ref{fig:diffdiagram}, employ a dual-encoder structure consisting of semantic and stochastic encoders. 
The semantic encoder focuses on preserving structural attributes, such as facial feature alignment, ensuring the morph retains the essential characteristics of both input identities. In contrast, the stochastic encoder captures fine-grained details, such as textures, hair direction, and clothing, enriching the morph’s visual fidelity while maintaining its identity relevance. Together, this dual-encoder structure provides a balance between maintaining overall structure and incorporating finer details. 
During the interpolation process, the semantic encoder streams are blended linearly to maintain consistent facial landmarks. Simultaneously, the stochastic encoder outputs are interpolated using spherical linear interpolation, allowing smooth and natural blending of fine-grained appearance features such as skin texture, hair style, and clothing patterns. This dual-path interpolation mechanism ensures the structural coherence and realistic textural details  
 of generated morphed.
To facilitate the effective blending of attributes, three fundamental functions are incorporated into the framework: 1) the image space preprocessing function $\xi$, 2) the image space interpolation function $\ell_\X$, and 3) the latent space interpolation function $\ell_\Z$.
The image space preprocessing function $\xi: \X \times \X \to \X$, aligns and enhances key features from the input images before encoding, ensuring consistency in structural attributes. The image space interpolation function, $\ell_\X(u, v; \gamma) = \gamma u + (1 - \gamma) v$, blends the two input images linearly, where $\gamma \in [0, 1]$ determines the blending ratio and controls the influence of each image. Finally, the latent space interpolation function, as defined in \cite{song2020denoising}, handles non-linear relationships in the latent space: 
\begin{equation}
\ell_\Z(u, v; \gamma) = \frac{\sin((1 - \gamma) \theta)}{\sin \theta}u + \frac{\sin(\gamma \theta)}{\sin \theta}v,
\end{equation}
where $\theta = \frac{\arccos(u \cdot v)}{|u|, |v|}$. This function enhances the realism of the generated morphs by addressing the complexity of interpolating stochastic codes and preserving semantic coherence.

While WaFusion builds on validated techniques such as diffusion autoencoders and wavelet transforms, its novelty lies in combining these domains for face morphing. By selectively applying generative processing only on low-frequency components while maintaining high-frequency details through efficient averaging, WaFusion uniquely balances computational efficiency and morph quality, which has not been explored in previous morphing methods. This hybrid approach results in high-quality, realistic face morphs that are robust against detection systems.

\section{Experiment and Results}
\subsection{Datasets}
We use four datasets: WVU Twin \footnote[2]{
The dataset is available upon request. To gain access, please contact Dr. Jeremy Dawson at \href{mailto:jeremy.dawson@mail.wvu.edu}{jeremy.dawson@mail.wvu.edu}.
} \cite{o2022identical}, FRLL \cite{debruine2017face}, FRGC \cite{phillips2005overview}, and FERET \cite{phillips1998feret}. WVU Twin contains frontal images with neutral expressions and plain backgrounds. It includes 2,268 unique identities, with some subjects appearing multiple times in collections spanning different years. This results in morph pairs with naturally high visual similarity.
The images have resolutions ranging from $2848 \times 4288$ to $5760 \times 3840$.

The other datasets contain passport-style frontal images under ideal lighting: FRLL (102 IDs, $413 \times 531$), FRGC (765 IDs subset, $1704 \times 2272$), and FERET (1,199 IDs, $512 \times 768$). 
We generate 2,971 WVU Twin-based, 1,122 FRLL-based, 964 FRGC-based, and 529 FERET-based morphs.

Baseline comparisons include FaceMorpher \cite{facemorpher}, OpenCV \cite{opencv}, StyleGAN \cite{karras2019style}, WebMorpher \cite{webmorph} (on FRLL only), and Diffusion Autoencoders \cite{Preechakul2021DiffusionAT, blasingame2024leveraging}. Morphing protocols follow Neubert \textit{et al.} for the AMSL dataset applied to FRLL \cite{neubert2018extended}, and Scherhag \textit{et al.} for FERET and FRGC \cite{scherhag2020deep}. We also include an LL-only baseline (LL-morphs), where only the low-frequency sub-band is morphed and high-frequency sub-bands are averaged.

\subsection{Implementations Details}
All the images are resized to $512 \times 512$, and aligned via facial landmarks. A single-level Haar wavelet transform is applied, resulting in four $256 \times 256$ sub-bands. The LL sub-band is input into the morph generator, while high-frequency sub-bands are averaged. The inverse wavelet transform reconstructs the final $512 \times 512$ morph.
Morph generation employs pre-trained Diffusion Autoencoders \cite{Preechakul2021DiffusionAT} using DDIM with 100 steps. All experiments are conducted on an NVIDIA Titan RTX GPU with 24 GB of VRAM. 

For verification, we use FaceNet \cite{schroff2015facenet}, a widely adopted framework for facial recognition, with an Inception backbone \cite{szegedy2015going}, pre-trained on the VGGFace2 \cite{cao2018vggface2}. FaceNet is leveraged to quantify look-alikes by generating compact feature embeddings for input images. It uses a triplet loss, where the Euclidean distance ($L_2$) for embeddings of the same identity are positive examples, and differing identities are considered negative examples \cite{schroff2015facenet}. 

\begin{figure*}[ht]
    \centering
    \begin{subfigure}[t]{0.24\textwidth}
        \centering
        \includegraphics[width=\textwidth]{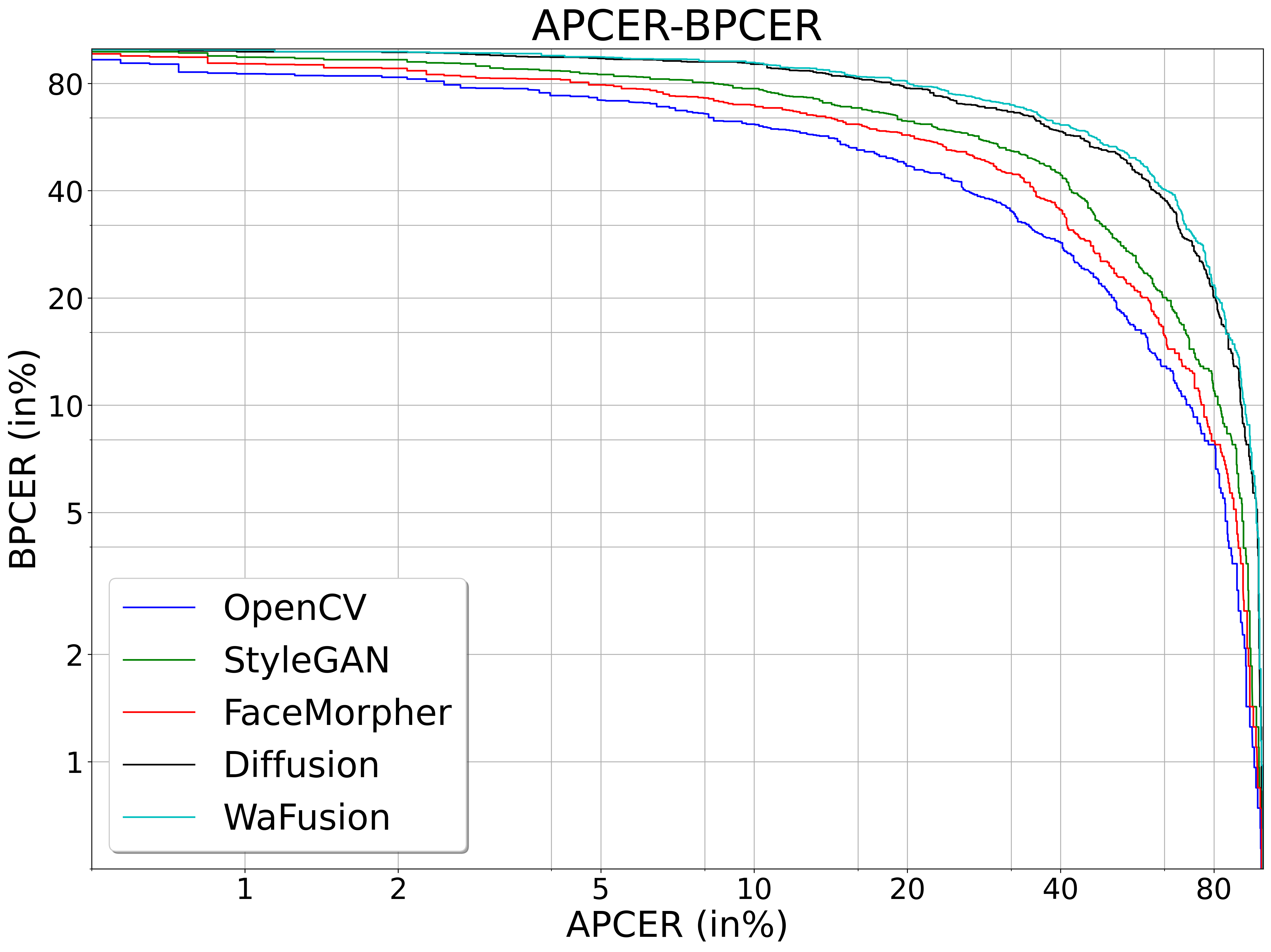}
        \caption{}
    \end{subfigure}
    \hfill
    \begin{subfigure}[t]{0.24\textwidth}
        \centering
        \includegraphics[width=\textwidth]{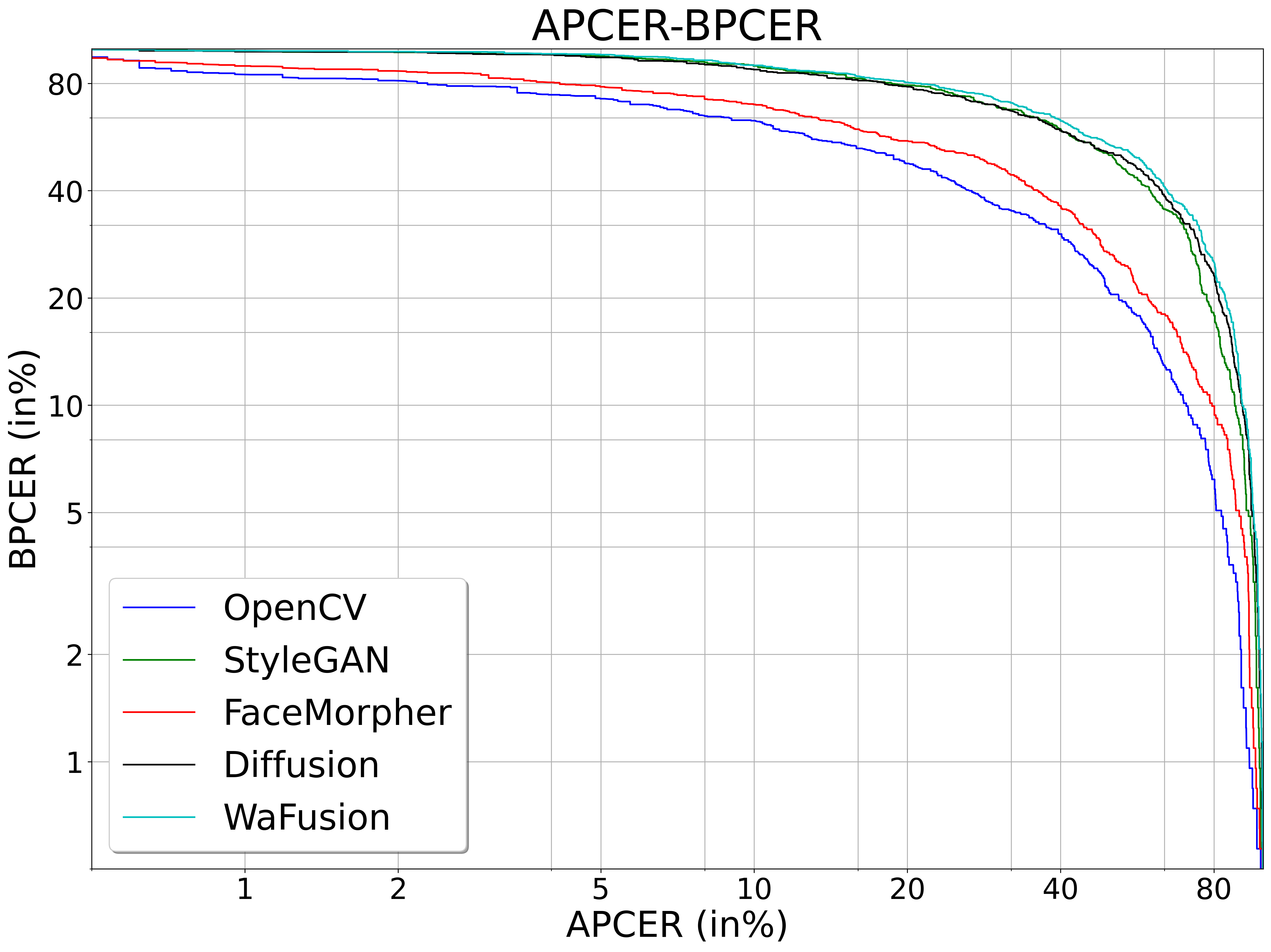}
        \caption{}
    \end{subfigure}
    \hfill
    \begin{subfigure}[t]{0.24\textwidth}
        \centering
        \includegraphics[width=\textwidth]{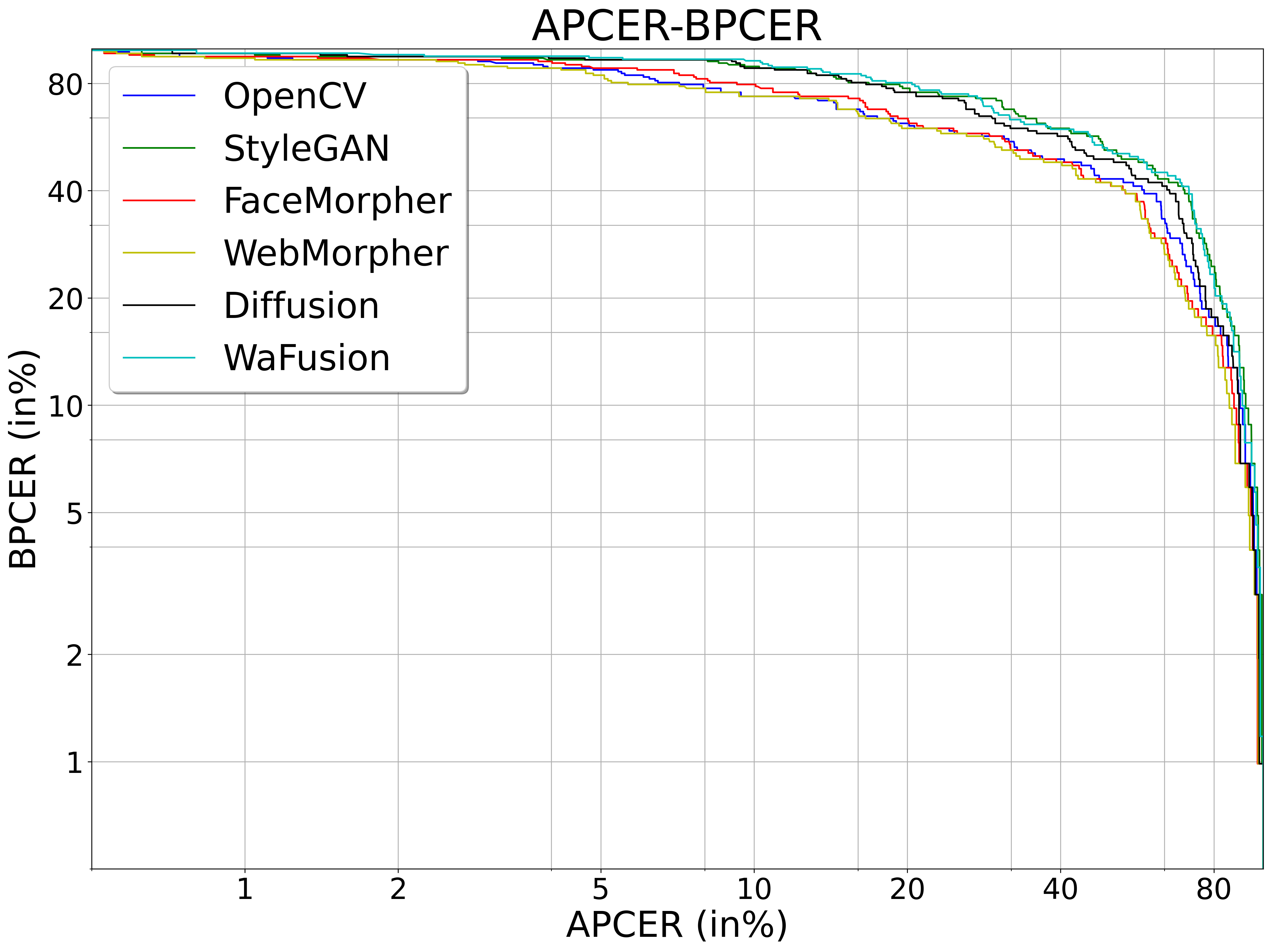}
        \caption{}
    \end{subfigure}
    \hfill
    \begin{subfigure}[t]{0.24\textwidth}
        \centering
        \includegraphics[width=\textwidth]{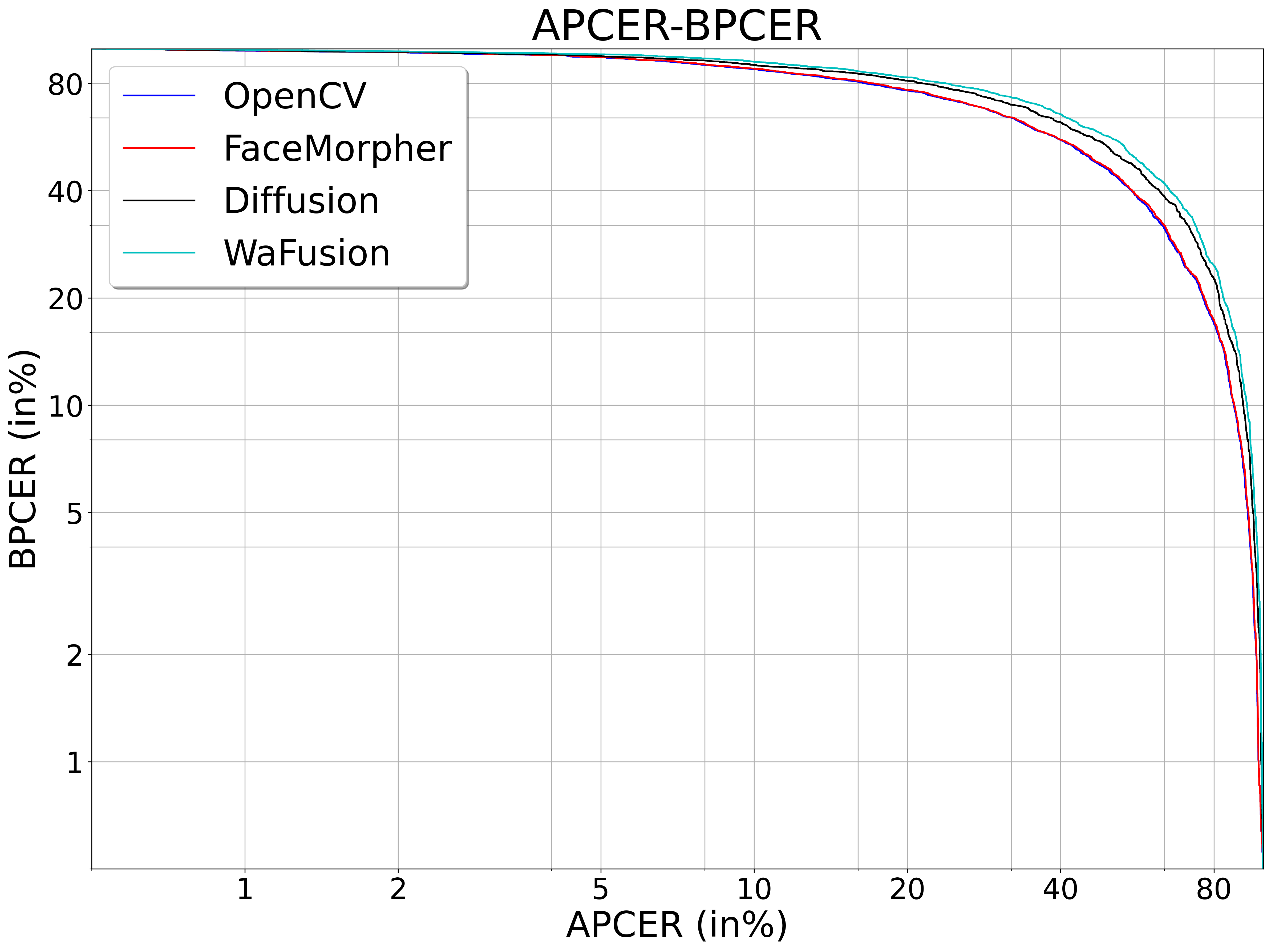}
        \caption{}
    \end{subfigure}

    \caption{APCER-BPCER curves for FaceNet verifier across (a) FERET, (b) FRGC, (c) FRLL, and (d) WVU Twin datasets.}
    \label{fig: ab_plots}
\end{figure*}
\subsection{Evaluation Metrics}
To evaluate the effectiveness of WaFusion, we use several key metrics: Area Under the Curve (AUC), Attack Presentation Classification Error Rate (APCER), Bona Fide Presentation Classification Error Rate (BPCER), and Equal Error Rate (EER). 
These metrics follow the ISO/IEC 30107-3 framework \cite{iso30107-3} and are commonly used in morphing attack detection.

The AUC is derived from the Receiver Operating Characteristic (ROC) curve, which plots the true positive rate against the false positive rate at various thresholds \cite{yang2022auc}. The EER represents the point where the FAR equals the False Rejection Rate (FRR) \cite{raja2020morphing}. APCER measures the ratio of morph attack samples incorrectly classified as bona fide presentations, while BPCER quantifies the percentage of genuine images misclassified as morphs \cite{ramachandra2017presentation, Venkatesh2020FaceMA}.


\begin{figure}[ht]
\begin{center}
    \includegraphics[width=\linewidth]{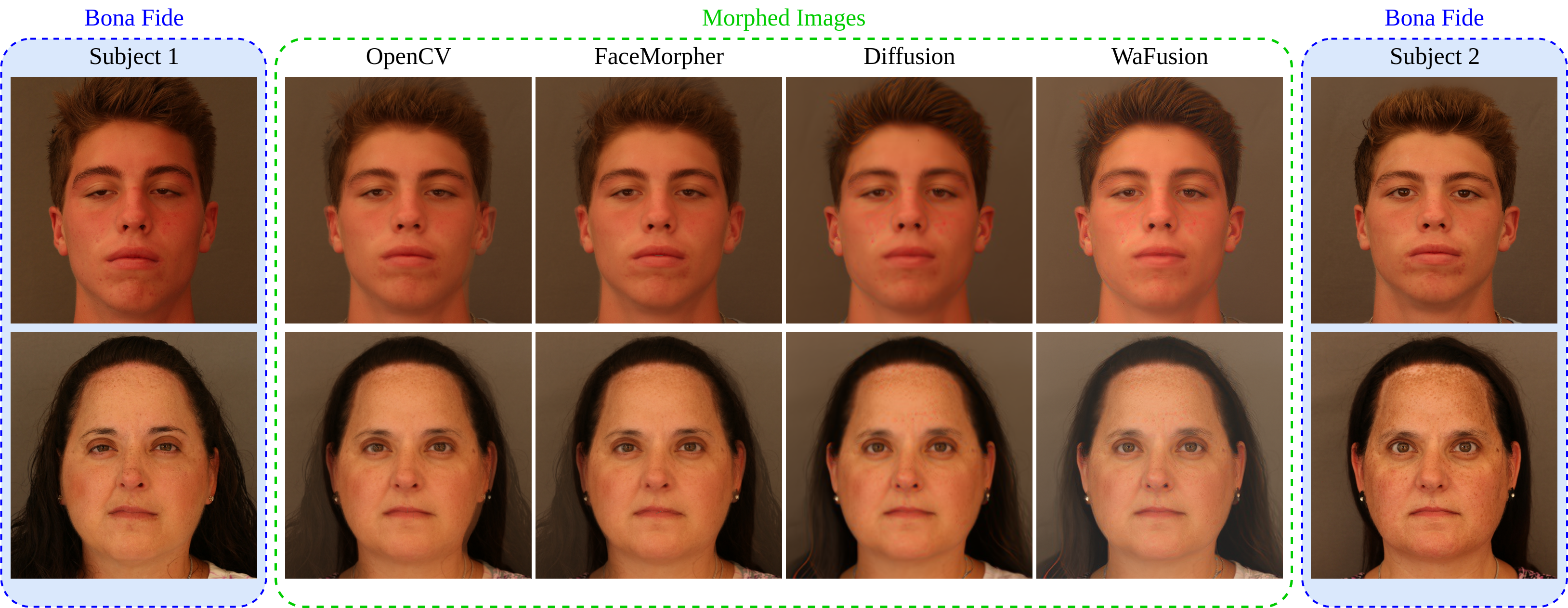}
\end{center}
   \caption{Visual comparison of morph images generated by OpenCV, FaceMorpher, Diffusion, and WaFusion using the WVU Twin dataset. Bona fide images are shown in blue boxes and morphs are shown in green boxes.}
\label{fig:visual_twin}
\end{figure}

\renewcommand{\arraystretch}{1.2}
\begin{table*}[ht!]
    \centering
    \caption{Performance of FaceNet verifier across both high-similarity (WVU Twin) and diverse-appearance datasets (FRLL, FERET, FRGC) evaluated using AUC, APCER, BPCER, and EER.}
    \resizebox{0.83\textwidth}{!}{%
    \begin{tabular}{|c|c|c|ccc|ccc|c|}
    \hline
    \multirow{2}{*}{\textbf{Datasets}} & \multirow{2}{*}{\textbf{Methods}} & \multirow{2}{*}{\textbf{AUC $\downarrow$}} & \multicolumn{3}{c|}{\textbf{APCER@BPCER (\%) $\uparrow$}}                                         & \multicolumn{3}{c|}{\textbf{BPCER@APCER(\%) $\uparrow$}}                                          & \multirow{2}{*}{\textbf{EER (\%)} $\uparrow$} \\ \cline{4-9}
                                       &                                   &                                    & \multicolumn{1}{c|}{\textbf{5\%}} & \multicolumn{1}{c|}{\textbf{10\%}} & \textbf{30\%} & \multicolumn{1}{c|}{\textbf{5\%}} & \multicolumn{1}{c|}{\textbf{10\%}} & \textbf{30\%} &                                    \\ \hline \hline
    \multirow{5}{*}{FRGC}              & OpenCV \cite{opencv}                           & 0.7267                             & \multicolumn{1}{c|}{71.294}       & \multicolumn{1}{c|}{61.351}        & 35.647        & \multicolumn{1}{c|}{84.631}       & \multicolumn{1}{c|}{71.339}        & 40.291        & 34.145                             \\ \cline{2-10} 
                                       & FaceMorpher \cite{facemorpher}                      & 0.6735                             & \multicolumn{1}{c|}{77.861}       & \multicolumn{1}{c|}{69.231}        & 46.341        & \multicolumn{1}{c|}{91.069}       & \multicolumn{1}{c|}{80.166}        & 47.04         & 37.523                             \\ \cline{2-10} 
                                       & StyleGAN \cite{karras2019style}                         & 0.5073                             & \multicolumn{1}{c|}{94.746}       & \multicolumn{1}{c|}{88.555}        & 68.292        & \multicolumn{1}{c|}{94.496}       & \multicolumn{1}{c|}{88.681}        & 71.131        & 50.281                             \\ \cline{2-10} 
                                       & Diffusion \cite{Preechakul2021DiffusionAT}                         & 0.4955                             & \multicolumn{1}{c|}{94.559}       & \multicolumn{1}{c|}{86.679}        & 68.292        & \multicolumn{1}{c|}{96.053}       & \multicolumn{1}{c|}{91.381}        & 73.624        & 50.656                             \\ \cline{2-10} 
                                       & \textbf{WaFusion}                          & \textbf{0.4843}                             & \multicolumn{1}{c|}{\textbf{95.717}}       & \multicolumn{1}{c|}{\textbf{89.453}}        & \textbf{71.422}        & \multicolumn{1}{c|}{\textbf{96.148}}       & \multicolumn{1}{c|}{\textbf{92.487}}        & \textbf{75.861}        & \textbf{52.593 }                            \\ \hline \hline
    \multirow{5}{*}{FERET}             & OpenCV \cite{opencv}                           & 0.7295                             & \multicolumn{1}{c|}{71.455}       & \multicolumn{1}{c|}{60.302}        & 36.862        & \multicolumn{1}{c|}{85.037}       & \multicolumn{1}{c|}{72.727}        & 37.31         & 32.703                             \\ \cline{2-10} 
                                       & FaceMorpher \cite{facemorpher}                      & 0.6798                             & \multicolumn{1}{c|}{77.315}       & \multicolumn{1}{c|}{68.241}        & 45.557        & \multicolumn{1}{c|}{88.825}       & \multicolumn{1}{c|}{76.515}        & 43.56         & 37.618                             \\ \cline{2-10} 
                                       & StyleGAN \cite{karras2019style}                         & 0.6295                             & \multicolumn{1}{c|}{83.742}       & \multicolumn{1}{c|}{76.37}         & 52.741        & \multicolumn{1}{c|}{91.477}       & \multicolumn{1}{c|}{82.765}        & 50.568        & 41.398                             \\ \cline{2-10} 
                                       & Diffusion \cite{Preechakul2021DiffusionAT}                         & 0.501                              & \multicolumn{1}{c|}{93.383}       & \multicolumn{1}{c|}{90.359}        & 67.296        & \multicolumn{1}{c|}{97.348}       & \multicolumn{1}{c|}{90.909}        & 69.696        & 50.85                              \\ \cline{2-10} 
                                       & \textbf{WaFusion}                          & \textbf{0.4923}                             & \multicolumn{1}{c|}{\textbf{94.424}}       & \multicolumn{1}{c|}{\textbf{90.276}}        & \textbf{70.473}        & \multicolumn{1}{c|}{\textbf{97.299}}       & \multicolumn{1}{c|}{\textbf{92.247}}        & \textbf{73.163}        & \textbf{51.856}                             \\ \hline \hline
    \multirow{6}{*}{FRLL}              & OpenCV \cite{opencv}                           & 0.5741                             & \multicolumn{1}{c|}{85.294}       & \multicolumn{1}{c|}{73.529}        & 56.862        & \multicolumn{1}{c|}{95.983}       & \multicolumn{1}{c|}{91.147}        & 65.573        & 46.078                             \\ \cline{2-10} 
                                       & FaceMorpher \cite{facemorpher}                      & 0.5848                             & \multicolumn{1}{c|}{88.235}       & \multicolumn{1}{c|}{77.45}         & 56.862        & \multicolumn{1}{c|}{95.331}       & \multicolumn{1}{c|}{88.615}        & 61.179        & 44.117                             \\ \cline{2-10} 
                                       & WebMorpher \cite{webmorph}                       & 0.6                                & \multicolumn{1}{c|}{80.392}       & \multicolumn{1}{c|}{73.529}        & 52.941        & \multicolumn{1}{c|}{94.098}       & \multicolumn{1}{c|}{86.803}        & 60.163        & 43.137                             \\ \cline{2-10} 
                                       & StyleGAN \cite{karras2019style}                         & \textbf{0.4793}                             & \multicolumn{1}{c|}{93.137}       & \multicolumn{1}{c|}{88.235}        & \textbf{71.568}        & \multicolumn{1}{c|}{\textbf{97.788}}       & \multicolumn{1}{c|}{\textbf{93.529}}        & 74.856        & \textbf{51.96}                              \\ \cline{2-10} 
                                       & Diffusion \cite{Preechakul2021DiffusionAT}                         & 0.5149                             & \multicolumn{1}{c|}{93.137}       & \multicolumn{1}{c|}{88.235}        & 61.764        & \multicolumn{1}{c|}{95.61}        & \multicolumn{1}{c|}{89.64}         & 70.676        & 49.019                             \\ \cline{2-10} 
                                       & \textbf{WaFusion}                          & 0.4909                             & \multicolumn{1}{c|}{\textbf{94.403}}       & \multicolumn{1}{c|}{\textbf{90.662}}        & 65.177        & \multicolumn{1}{c|}{97.496}       & \multicolumn{1}{c|}{91.891}        & \textbf{75.891 }       & 50.798                             \\ \hline \hline
    \multirow{5}{*}{WVU Twin}          & OpenCV \cite{opencv}                           & 0.5349                             & \multicolumn{1}{c|}{93.886}       & \multicolumn{1}{c|}{86.876}        & 65.342        & \multicolumn{1}{c|}{93.797}       & \multicolumn{1}{c|}{88.246}        & 65.197        & 47.507                             \\ \cline{2-10} 
                                       & FaceMorpher \cite{facemorpher}                      & 0.5315                             & \multicolumn{1}{c|}{93.964}       & \multicolumn{1}{c|}{87.5}          & 65.537        & \multicolumn{1}{c|}{94.001}       & \multicolumn{1}{c|}{88.517}        & 65.678        & 47.507                             \\ \cline{2-10} 
                                       & Diffusion \cite{Preechakul2021DiffusionAT}                       & 0.4829                             & \multicolumn{1}{c|}{94.859}       & \multicolumn{1}{c|}{89.369}        & 71.495        & \multicolumn{1}{c|}{95.847}       & \multicolumn{1}{c|}{91.729}        & 73.335        & 50.506                             \\ \cline{2-10} 
                                       & \textbf{WaFusion}                          & \textbf{0.4715}                             & \multicolumn{1}{c|}{\textbf{96.296}}       & \multicolumn{1}{c|}{\textbf{91.583}}        & \textbf{73.983}        & \multicolumn{1}{c|}{\textbf{96.934}}       & \multicolumn{1}{c|}{\textbf{93.274}}        & \textbf{75.218}        & \textbf{52.681}                             \\ \hline
    \end{tabular}%
    }
    \label{tab: metrics}
\end{table*}
\subsection{Vulnerability Analysis}

This section evaluates the proposed WaFusion framework using the FRGC, FERET, FRLL, and WVU Twin datasets. The WVU Twin dataset provides controlled and high-similarity morphs. To further evaluate across broader appearance variations and conditions, we use FRLL, FERET, and FRGC. This multi-dataset evaluation ensures generalization beyond highly similar subjects. The evaluation includes visual comparisons, quantitative metrics, and an ablation study to assess individual component contributions. 

A visual comparison of morph generation methods is presented in Fig. \ref{fig:visual_twin}, using the WVU Twin dataset. WaFusion generates morphs at $512 \times 512$, surpassing the $256 \times 256$ resolution of the Diffusion approach. This demonstrates WaFusion's efficiency in achieving higher-quality morphs without increasing computational costs. Additionally, WaFusion benefits from targeting the LL sub-band, which captures most of the structural content, while averaging the high-frequency sub-bands to reduce unnecessary computation.

WaFusion morphs exhibit superior realism and detail, particularly in finer features such as facial textures and hair. In contrast, Diffusion morphs lack resolution, and traditional methods like OpenCV and FaceMorpher produce visible artifacts in facial contours and hairlines. The WVU Twin dataset, characterized by highly similar identities, poses challenges for all methods, yet WaFusion consistently minimizes artifacts and delivers visually consistent results. This highlights the framework's robustness even under extreme inter-subject similarity.

\begin{figure*}[hb]
    \centering
    \begin{subfigure}[t]{0.24\textwidth}
        \centering
        \includegraphics[width=\textwidth]{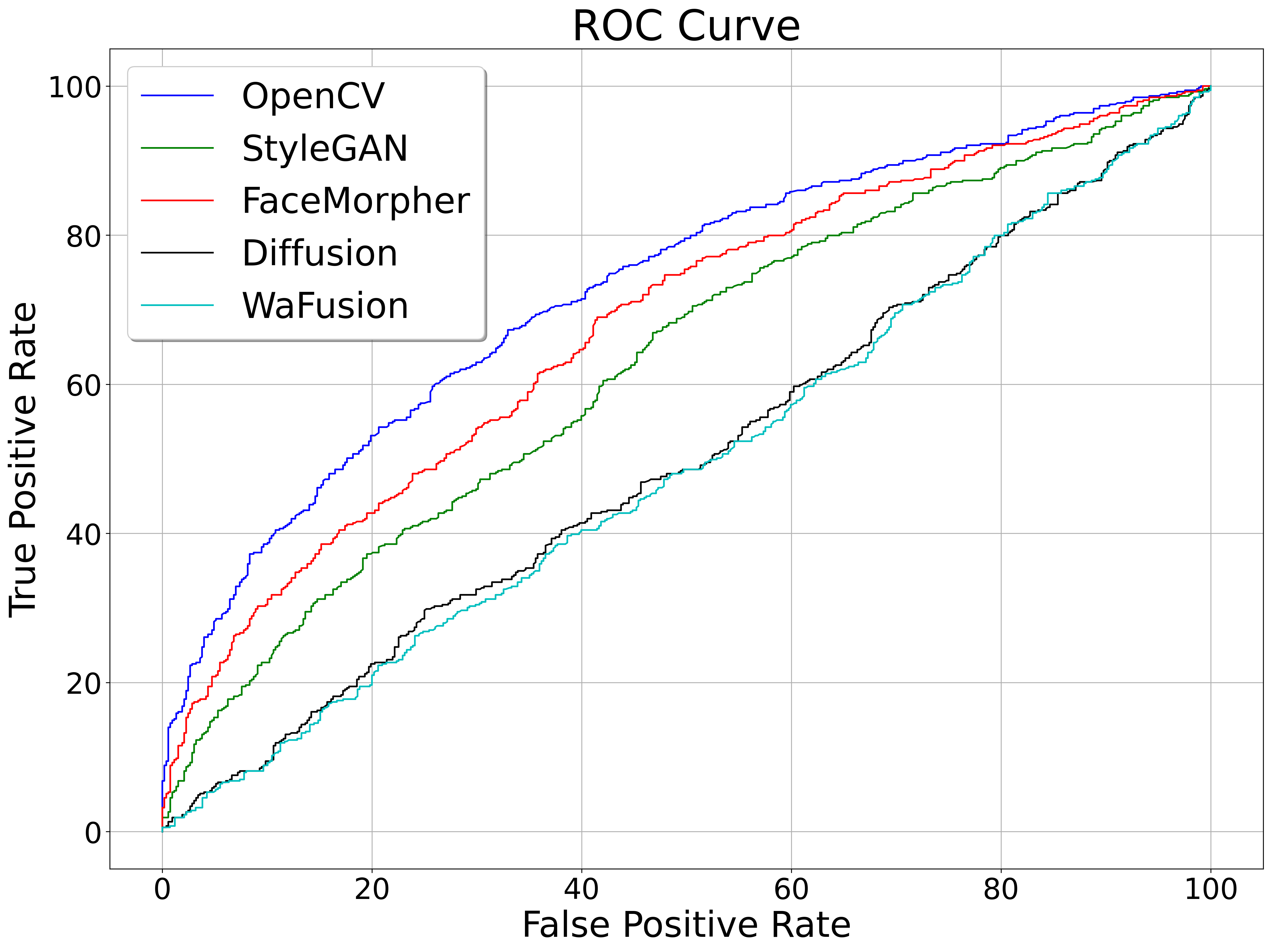}
        \caption{}
    \end{subfigure}
    \hfill
    \begin{subfigure}[t]{0.24\textwidth}
        \centering
        \includegraphics[width=\textwidth]{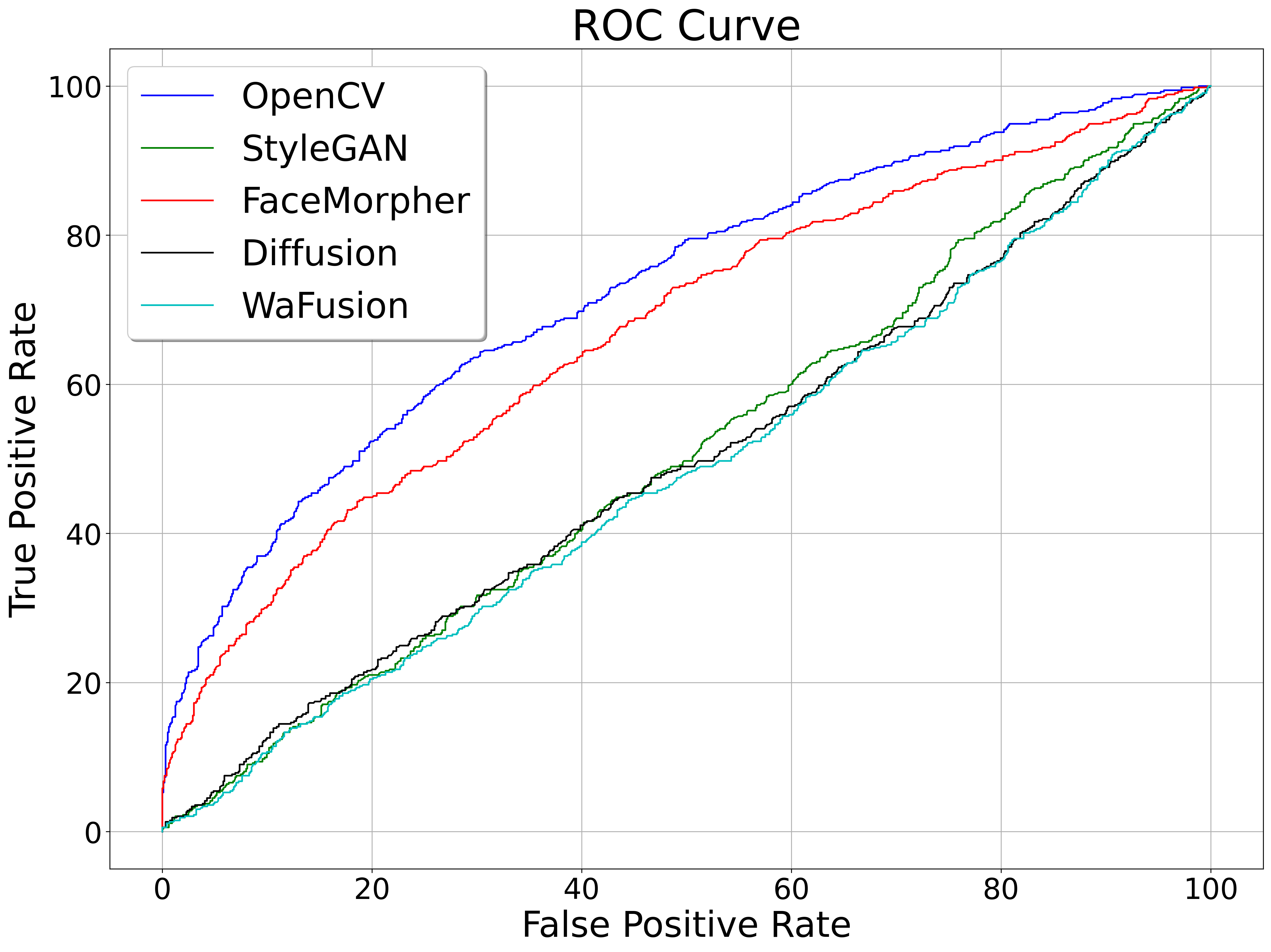}
        \caption{}
    \end{subfigure}
    \hfill
    \begin{subfigure}[t]{0.24\textwidth}
        \centering
        \includegraphics[width=\textwidth]{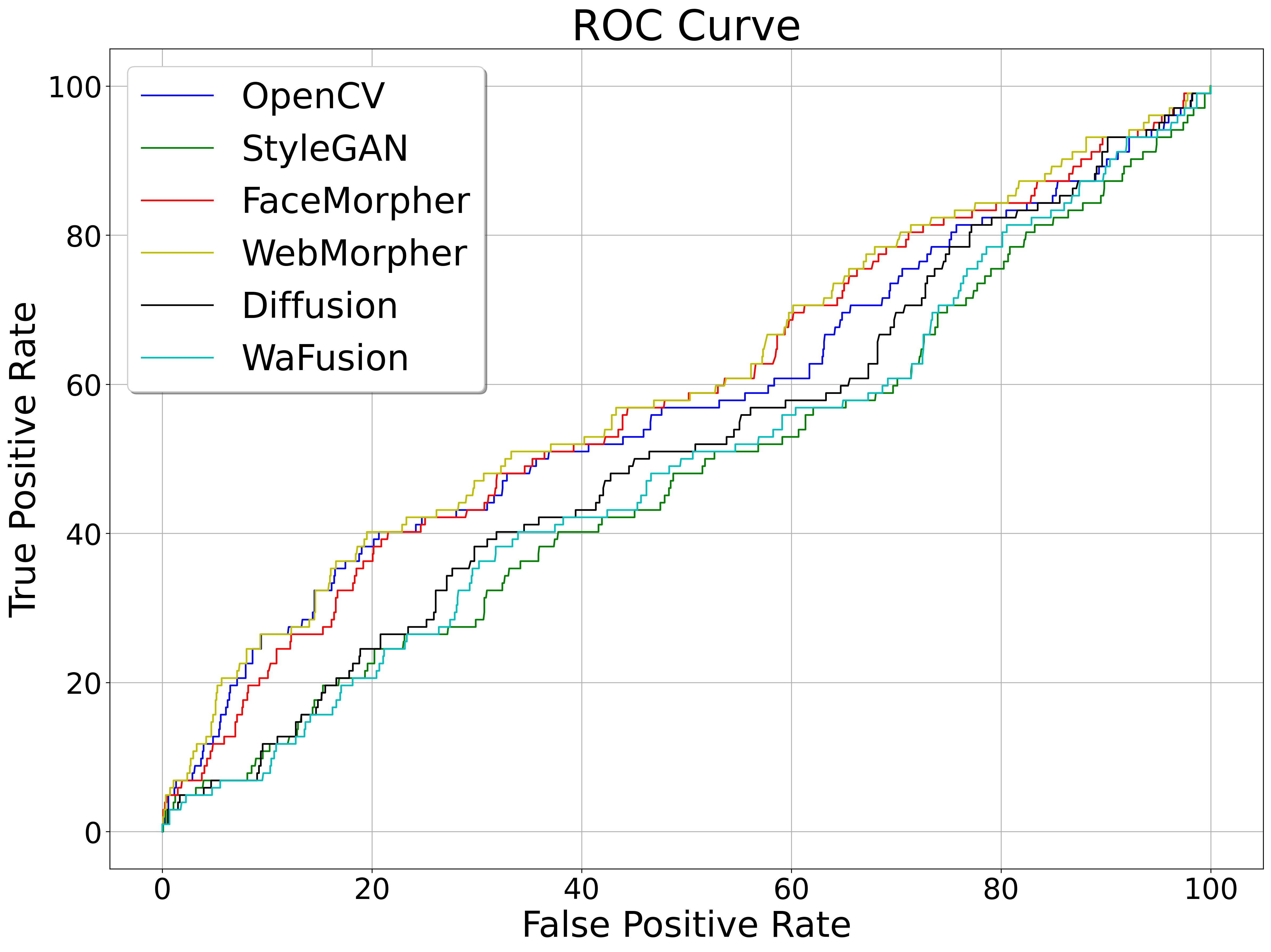}
        \caption{}
    \end{subfigure}
    \hfill
    \begin{subfigure}[t]{0.24\textwidth}
        \centering
        \includegraphics[width=\textwidth]{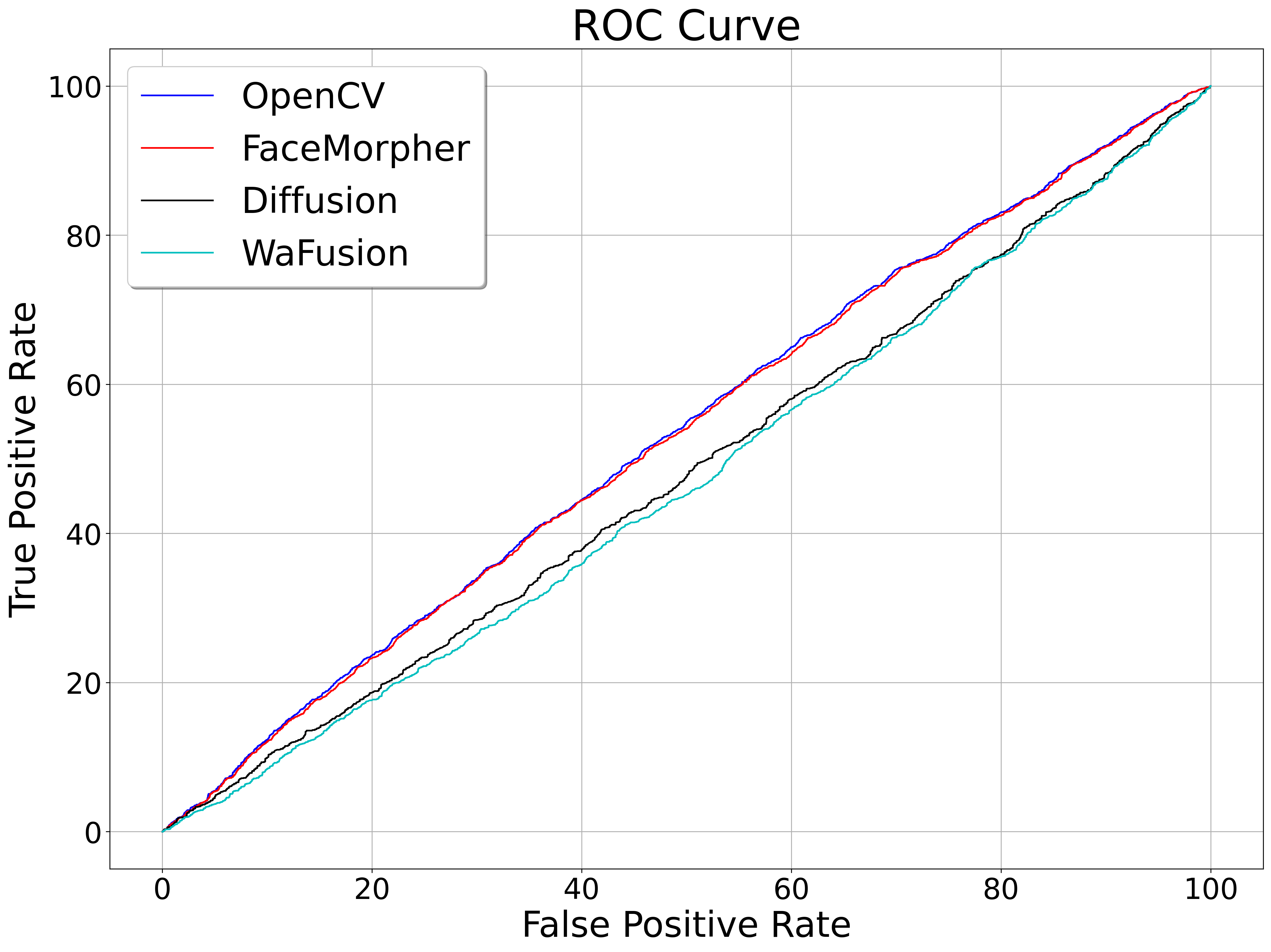}
        \caption{}
    \end{subfigure}

    \caption{ROC curves for FaceNet verifier across (a) FERET, (b) FRGC, (c) FRLL, and (d) WVU Twin datasets.}
    \label{fig: roc_plots}
\end{figure*}

The metrics in Table \ref{tab: metrics} provide a quantitative comparison of WaFusion with OpenCV, FaceMorpher, StyleGAN, and Diffusion across various datasets, using AUC, APCER, BPCER, and EER. WaFusion stands out with the lowest EER and competitive results across most datasets, highlighting its effectiveness in generating realistic and diverse morphs. On the FRGC and FERET, WaFusion outperforms other methods, achieving high-quality morphs with minimal artifacts. On the FRLL, StyleGAN slightly outperforms WaFusion in some metrics due to the dataset’s limited diversity, but WaFusion’s morphs remain visually more realistic. For WVU Twin, WaFusion achieves the lowest AUC and EER, demonstrating its effectiveness even with high-similarity identities.

Fig. \ref{fig: ab_plots} and Fig. \ref{fig: roc_plots} illustrate the APCER-BPCER and ROC curves, respectively, using the FaceNet, further validating WaFusion's effectiveness. These curves demonstrate our method's ability to generate more challenging morphs for verifiers, particularly at stricter thresholds, achieving a favorable balance between attack success rates and bona fide misclassification rates across all datasets.

\begin{figure*}[t]
\begin{center}
    \includegraphics[width=0.8\linewidth]{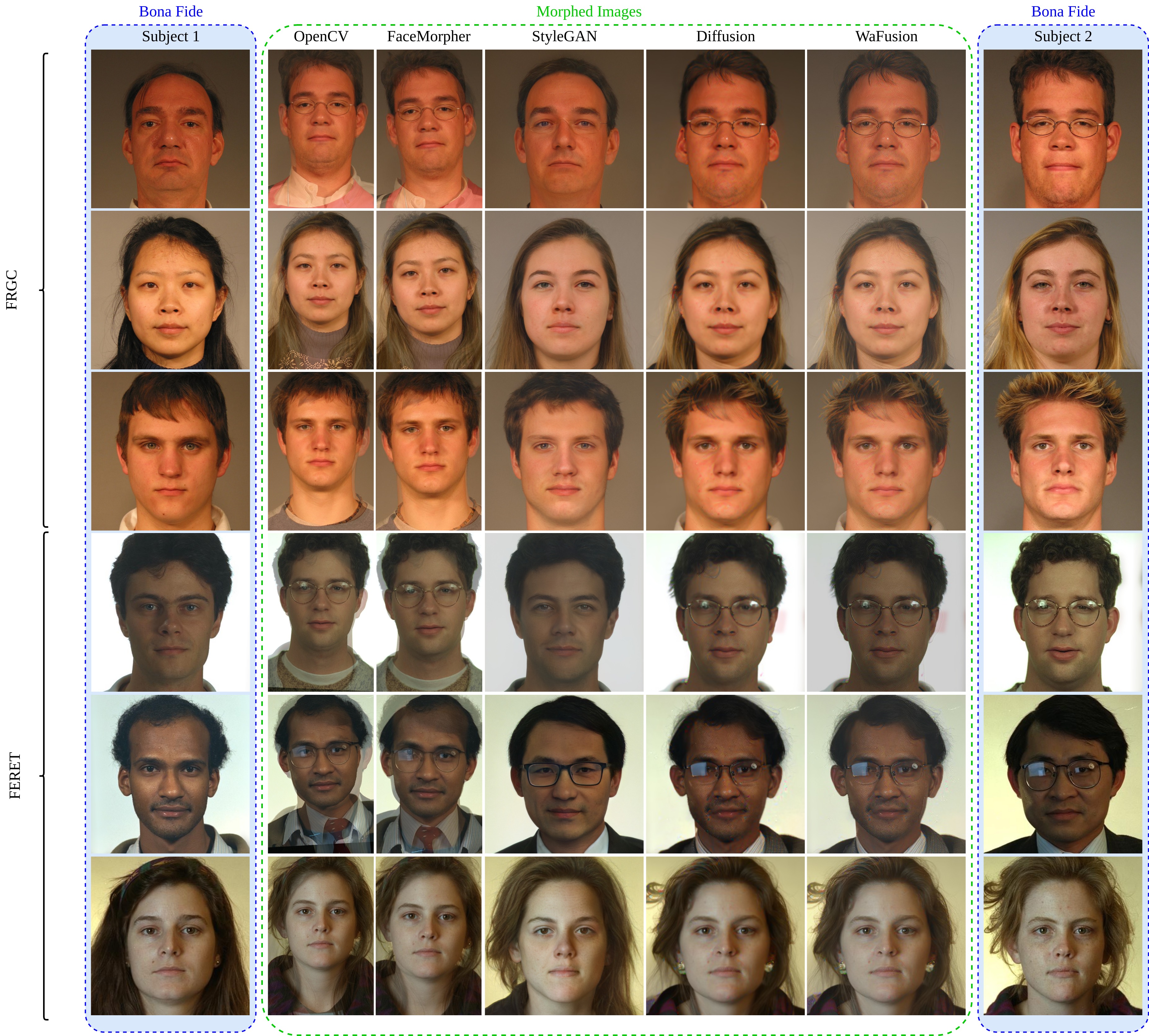}

   \caption{Morph images generated by different methods on the FRGC and FERET datasets, showing bona fide images (blue boxes) and morphs (green boxes).}
\label{fig:visual_comparison}
\end{center}
\end{figure*}

A broader visual comparison, shown in Fig. \ref{fig:visual_comparison}, highlights WaFusion’s superior realism and consistency across morphs generated for the FRGC and FERET. While StyleGAN produces high-quality images, its morphs lack realism due to GAN-specific limitations such as mode collapse. Diffusion-based morphs suffer from resolution constraints, while traditional methods like OpenCV and FaceMorpher generate artifacts in facial features. WaFusion consistently delivers detailed, realistic morphs, reinforcing its adaptability and scalability across datasets with varying characteristics.

\subsection{Ablation Study}

The ablation study in Table \ref{tab: ablation} compares two variations of WaFusion: morphing all four sub-bands (LL, LH, HL, HH) versus morphing only the LL sub-band while averaging the high-frequency sub-bands. Results using Learned Perceptual Image Patch Similarity (LPIPS) \cite{zhang2018unreasonable} and Structural Similarity Index Measure (SSIM) \cite{wang2004image} show that LL-only morphing achieves comparable performance, as the LL sub-band captures most structural information, while high-frequency sub-bands mainly represent edge details.

\begin{table}[h!]
    \caption{Comparison of morph generation methods on the Twin dataset, using LPIPS and SSIM metrics.}
    \centering
    \scriptsize
    \resizebox{0.36\textwidth}{!}{%
    \begin{tabular}{|c|c|c|}
        \hline
        \textbf{Method} & \textbf{LPIPS $\downarrow$} & \textbf{SSIM $\uparrow$} \\
        \hline
        \textbf{WaFusion} (All sub-bands)  & 0.243 & 0.792 \\
        \textbf{WaFusion} (LL sub-band) & \textbf{0.239} & \textbf{0.798} \\
        \hline
    \end{tabular}%
    }
    \label{tab: ablation}
\end{table}

Moreover, processing all four sub-bands requires approximately four times more computation than morphing, confirming the design's efficiency without compromising morph quality.

\subsection{Future Work}

Future work will focus on improving the scalability, generalization, and performance of WaFusion. While this study uses a single-level Haar wavelet for simplicity, future versions will explore multi-level decompositions and alternative bases (e.g., Daubechies) to capture fine-grained morphological details. We will also investigate selective processing of high-frequency sub-bands at higher wavelet levels to better balance morph quality and computation. Another direction involves extending WaFusion to real-time scenarios, such as video-based biometric authentication. These enhancements will further position WaFusion as a robust and adaptable framework for biometric security. 

\section{Conclusion}

In this study, we introduce WaFusion, a hybrid framework combining wavelet decomposition and diffusion models to generate high-quality facial morphs. WaFusion preserves structural integrity and fine texture while minimizing perceptual artifacts, all without additional computational cost.
Extensive evaluations on FRGC, FERET, FRLL, and WVU Twin demonstrate WaFusion's superiority over state-of-the-art methods across biometric metrics including APCER, BPCER, and EER. Its ability to produce realistic and challenging morphs underscores its potential to advance morph generation and strengthen biometric security systems.

{\small
\bibliographystyle{ieee}
\bibliography{egbib}

\begin{thebibliography}{10}\itemsep=-1pt

\bibitem{ashwini2023generation}
K.~Ashwini, D.~Nagajyothi, C.~Ramakrishna, and V.~Jyothi.
\newblock Generation of high-quality realistic faces with {StyleGAN}.
\newblock In {\em 2023 4th IEEE Global Conference for Advancement in Technology (GCAT)}, pages 1--7. IEEE, 2023.

\bibitem{blasingame2024leveraging}
Z.~Blasingame and C.~Liu.
\newblock Leveraging diffusion for strong and high quality face morphing attacks.
\newblock {\em IEEE Transactions on Biometrics, Behavior, and Identity Science}, 2024.

\bibitem{blasingame2024fast}
Z.~W. Blasingame and C.~Liu.
\newblock Fast-dim: Towards fast diffusion morphs.
\newblock {\em IEEE Security \& Privacy}, 2024.

\bibitem{bowyer2004face}
K.~W. Bowyer.
\newblock Face recognition technology: security versus privacy.
\newblock {\em IEEE Technology and society magazine}, 23(1):9--19, 2004.

\bibitem{cao2018vggface2}
Q.~Cao, L.~Shen, W.~Xie, O.~M. Parkhi, and A.~Zisserman.
\newblock Vggface2: A dataset for recognising faces across pose and age.
\newblock In {\em 2018 13th IEEE international conference on automatic face \& gesture recognition (FG 2018)}, pages 67--74. IEEE, 2018.

\bibitem{cootes2001active}
T.~F. Cootes, G.~J. Edwards, and C.~J. Taylor.
\newblock Active appearance models.
\newblock {\em IEEE Transactions on pattern analysis and machine intelligence}, 23(6):681--685, 2001.

\bibitem{damer2021regenmorph}
N.~Damer, K.~Raja, M.~S{\"u}{\ss}milch, S.~Venkatesh, F.~Boutros, M.~Fang, F.~Kirchbuchner, R.~Ramachandra, and A.~Kuijper.
\newblock Regenmorph: Visibly realistic {GAN} generated face morphing attacks by attack re-generation.
\newblock In {\em Advances in Visual Computing: 16th International Symposium, ISVC 2021, Virtual Event, October 4-6, 2021, Proceedings, Part I}, pages 251--264. Springer, 2021.

\bibitem{damer2018morgan}
N.~Damer, A.~M. Saladie, A.~Braun, and A.~Kuijper.
\newblock {MorGAN}: Recognition vulnerability and attack detectability of face morphing attacks created by generative adversarial network.
\newblock In {\em 2018 IEEE 9th international conference on biometrics theory, applications and systems (BTAS)}, pages 1--10. IEEE, 2018.

\bibitem{daubechies1992ten}
I.~Daubechies.
\newblock {\em Ten lectures on wavelets}.
\newblock SIAM, 1992.

\bibitem{webmorph}
L.~DeBruine.
\newblock debruine/webmorph: Beta release 2.
\newblock {\em Zenodo https://doi. org/10}, 5281, 2018.

\bibitem{debruine2017face}
L.~DeBruine and B.~Jones.
\newblock Face research lab london set.
\newblock {\em Psychol. Methodol. Des. Anal}, 3, 2017.

\bibitem{dhariwal2021diffusion}
P.~Dhariwal and A.~Nichol.
\newblock Diffusion models beat {GANs} on image synthesis.
\newblock {\em Advances in neural information processing systems}, 34:8780--8794, 2021.

\bibitem{Ferrara2014TheMP}
M.~Ferrara, A.~Franco, and D.~Maltoni.
\newblock The magic passport.
\newblock {\em IEEE International Joint Conference on Biometrics}, pages 1--7, 2014.

\bibitem{ferrara2019decoupling}
M.~Ferrara, A.~Franco, and D.~Maltoni.
\newblock Decoupling texture blending and shape warping in face morphing.
\newblock In {\em 2019 international conference of the biometrics special interest group (BIOSIG)}, pages 1--5. IEEE, 2019.

\bibitem{iso30107-3}
I.~O. for Standardization.
\newblock {ISO/IEC 30107-3:2023: Information technology -- Biometric presentation attack detection}.
\newblock Technical report, ISO, 2023.

\bibitem{godage2022analyzing}
S.~R. Godage, F.~L{\o}v{\aa}sdal, S.~Venkatesh, K.~Raja, R.~Ramachandra, and C.~Busch.
\newblock Analyzing human observer ability in morphing attack detection—where do we stand?
\newblock {\em IEEE Transactions on Technology and Society}, 4(2):125--145, 2022.

\bibitem{GomezBarrero2022OnTD}
M.~Gomez-Barrero, K.~B. Raja, C.~Rathgeb, A.~F. Sequeira, M.~Todisco, L.~Colbois, and S.~Marcel.
\newblock On the detection of morphing attacks generated by {GANs}.
\newblock {\em 2022 International Conference of the Biometrics Special Interest Group (BIOSIG)}, pages 1--5, 2022.

\bibitem{goodfellow2020generative}
I.~Goodfellow, J.~Pouget-Abadie, M.~Mirza, B.~Xu, D.~Warde-Farley, S.~Ozair, A.~Courville, and Y.~Bengio.
\newblock Generative adversarial networks.
\newblock {\em Communications of the ACM}, 63(11):139--144, 2020.

\bibitem{hamza2022comprehensive}
M.~Hamza, S.~Tehsin, M.~Humayun, M.~F. Almufareh, and M.~Alfayad.
\newblock A comprehensive review of face morph generation and detection of fraudulent identities.
\newblock {\em Applied Sciences}, 12(24):12545, 2022.

\bibitem{ho2020denoising}
J.~Ho, A.~Jain, and P.~Abbeel.
\newblock Denoising diffusion probabilistic models.
\newblock {\em Advances in neural information processing systems}, 33:6840--6851, 2020.

\bibitem{huang2023wavedm}
Y.~Huang, J.~Huang, J.~Liu, Y.~Dong, J.~Lv, and S.~Chen.
\newblock Wavedm: Wavelet-based diffusion models for image restoration.
\newblock {\em arXiv preprint arXiv:2305.13819}, 2023.

\bibitem{icao20159303}
D.~ICAO.
\newblock 9303-machine readable travel documents-part 9: Deployment of biometric identification and electronic storage of data in emrtds.
\newblock {\em International Civil Aviation Organization (ICAO)}, 123, 2015.

\bibitem{ivanovska2023face}
M.~Ivanovska and V.~{\v{S}}truc.
\newblock Face morphing attack detection with denoising diffusion probabilistic models.
\newblock In {\em 2023 11th International Workshop on Biometrics and Forensics (IWBF)}, pages 1--6. IEEE, 2023.

\bibitem{karras2019style}
T.~Karras, S.~Laine, and T.~Aila.
\newblock A style-based generator architecture for generative adversarial networks.
\newblock In {\em Proceedings of the IEEE/CVF conference on computer vision and pattern recognition}, pages 4401--4410, 2019.

\bibitem{kessler2024towards}
R.~Kessler, K.~Raja, J.~Tapia, and C.~Busch.
\newblock Towards minimizing efforts for morphing attacks—deep embeddings for morphing pair selection and improved morphing attack detection.
\newblock {\em Plos one}, 19(5):e0304610, 2024.

\bibitem{kim2024diffusion}
J.~Kim, C.~Oh, H.~Do, S.~Kim, and K.~Sohn.
\newblock Diffusion-driven gan inversion for multi-modal face image generation.
\newblock In {\em Proceedings of the IEEE/CVF Conference on Computer Vision and Pattern Recognition}, pages 10403--10412, 2024.

\bibitem{pywt}
G.~R. Lee, R.~Gommers, F.~Waselewski, K.~Wohlfahrt, and A.~O'Leary.
\newblock Pywavelets: A python package for wavelet analysis.
\newblock {\em Journal of Open Source Software}, 4(36):1237, 2019.

\bibitem{makrushin2020simulation}
A.~Makrushin, D.~Siegel, and J.~Dittmann.
\newblock Simulation of border control in an ongoing web-based experiment for estimating morphing detection performance of humans.
\newblock In {\em Proceedings of the 2020 ACM Workshop on Information Hiding and Multimedia Security}, pages 91--96, 2020.

\bibitem{mallat1989theory}
S.~Mallat.
\newblock A theory for multiresolution signal decomposition: The wavelet representation.
\newblock {\em IEEE Trans. Pattern Anal. Mach. Intell.}, 11:674--693, 1989.

\bibitem{opencv}
S.~Mallick.
\newblock Face morph using opencv—c++/python.
\newblock {\em LearnOpenCV}, 1(1), 2016.

\bibitem{neubert2018extended}
T.~Neubert, A.~Makrushin, M.~Hildebrandt, C.~Kraetzer, and J.~Dittmann.
\newblock Extended stirtrace benchmarking of biometric and forensic qualities of morphed face images.
\newblock {\em Iet Biometrics}, 7(4):325--332, 2018.

\bibitem{nichol2021improved}
A.~Q. Nichol and P.~Dhariwal.
\newblock Improved denoising diffusion probabilistic models.
\newblock In {\em International conference on machine learning}, pages 8162--8171. PMLR, 2021.

\bibitem{OHaire2021AdversariallyPW}
K.~O'Haire, S.~Soleymani, B.~Chaudhary, P.~Aghdaie, J.~M. Dawson, and N.~M. Nasrabadi.
\newblock Adversarially perturbed wavelet-based morphed face generation.
\newblock {\em 2021 16th IEEE International Conference on Automatic Face and Gesture Recognition (FG 2021)}, pages 01--05, 2021.

\bibitem{o2022identical}
K.~O'Haire, S.~Soleymani, B.~Chaudhary, J.~Dawson, and N.~M. Nasrabadi.
\newblock Identical twins face morph database generation.
\newblock In {\em 2022 IEEE International Joint Conference on Biometrics (IJCB)}, pages 1--9. IEEE, 2022.

\bibitem{phillips2005overview}
P.~J. Phillips, P.~J. Flynn, T.~Scruggs, K.~W. Bowyer, J.~Chang, K.~Hoffman, J.~Marques, J.~Min, and W.~Worek.
\newblock Overview of the face recognition grand challenge.
\newblock In {\em 2005 IEEE computer society conference on computer vision and pattern recognition (CVPR'05)}, volume~1, pages 947--954. IEEE, 2005.

\bibitem{phillips1998feret}
P.~J. Phillips, H.~Wechsler, J.~Huang, and P.~J. Rauss.
\newblock The feret database and evaluation procedure for face-recognition algorithms.
\newblock {\em Image and vision computing}, 16(5):295--306, 1998.

\bibitem{phung2023wavelet}
H.~Phung, Q.~Dao, and A.~Tran.
\newblock Wavelet diffusion models are fast and scalable image generators.
\newblock In {\em Proceedings of the IEEE/CVF Conference on Computer Vision and Pattern Recognition}, pages 10199--10208, 2023.

\bibitem{Preechakul2021DiffusionAT}
K.~Preechakul, N.~Chatthee, S.~Wizadwongsa, and S.~Suwajanakorn.
\newblock Diffusion autoencoders: Toward a meaningful and decodable representation.
\newblock {\em 2022 IEEE/CVF Conference on Computer Vision and Pattern Recognition (CVPR)}, pages 10609--10619, 2021.

\bibitem{price2022landmark}
S.~Price, S.~Soleymani, and N.~M. Nasrabadi.
\newblock Landmark enforcement and style manipulation for generative morphing.
\newblock In {\em 2022 IEEE International Joint Conference on Biometrics (IJCB)}, pages 1--10. IEEE, 2022.

\bibitem{facemorpher}
A.~Quek.
\newblock Facemorpher, 2019.

\bibitem{Raghavendra2016DetectingMF}
R.~Raghavendra, K.~B. Raja, and C.~Busch.
\newblock Detecting morphed face images.
\newblock {\em 2016 IEEE 8th International Conference on Biometrics Theory, Applications and Systems (BTAS)}, pages 1--7, 2016.

\bibitem{raja2020morphing}
K.~Raja, M.~Ferrara, A.~Franco, L.~Spreeuwers, I.~Batskos, F.~De~Wit, M.~Gomez-Barrero, U.~Scherhag, D.~Fischer, S.~K. Venkatesh, et~al.
\newblock Morphing attack detection-database, evaluation platform, and benchmarking.
\newblock {\em IEEE transactions on information forensics and security}, 16:4336--4351, 2020.

\bibitem{ramachandra2017presentation}
R.~Ramachandra and C.~Busch.
\newblock Presentation attack detection methods for face recognition systems: A comprehensive survey.
\newblock {\em ACM Computing Surveys (CSUR)}, 50(1):1--37, 2017.

\bibitem{ramachandra2024multispectral}
R.~Ramachandra, S.~Venkatesh, N.~Damer, N.~Vetrekar, and R.~S. Gad.
\newblock Multispectral imaging for differential face morphing attack detection: A preliminary study.
\newblock In {\em Proceedings of the IEEE/CVF Winter Conference on Applications of Computer Vision}, pages 6185--6193, 2024.

\bibitem{robertson2017fraudulent}
D.~J. Robertson, R.~S. Kramer, and A.~M. Burton.
\newblock Fraudulent {ID} using face morphs: Experiments on human and automatic recognition.
\newblock {\em PLoS One}, 12(3):e0173319, 2017.

\bibitem{roich2022pivotal}
D.~Roich, R.~Mokady, A.~H. Bermano, and D.~Cohen-Or.
\newblock Pivotal tuning for latent-based editing of real images.
\newblock {\em ACM Transactions on graphics (TOG)}, 42(1):1--13, 2022.

\bibitem{rombach2022high}
R.~Rombach, A.~Blattmann, D.~Lorenz, P.~Esser, and B.~Ommer.
\newblock High-resolution image synthesis with latent diffusion models.
\newblock In {\em Proceedings of the IEEE/CVF conference on computer vision and pattern recognition}, pages 10684--10695, 2022.

\bibitem{ronneberger2015u}
O.~Ronneberger, P.~Fischer, and T.~Brox.
\newblock {U-Net}: Convolutional networks for biomedical image segmentation.
\newblock In {\em Medical image computing and computer-assisted intervention--MICCAI 2015: 18th international conference, Munich, Germany, October 5-9, 2015, proceedings, part III 18}, pages 234--241. Springer, 2015.

\bibitem{saadabadi2024boosting}
M.~S.~E. Saadabadi, S.~R. Malakshan, S.~R. Hosseini, and N.~M. Nasrabadi.
\newblock Boosting unconstrained face recognition with targeted style adversary.
\newblock In {\em 2024 IEEE International Joint Conference on Biometrics (IJCB)}, pages 1--11. IEEE, 2024.

\bibitem{sarkar2020vulnerability}
E.~Sarkar, P.~Korshunov, L.~Colbois, and S.~Marcel.
\newblock Vulnerability analysis of face morphing attacks from landmarks and generative adversarial networks.
\newblock {\em arXiv preprint arXiv:2012.05344}, 2020.

\bibitem{sarkar2022gan}
E.~Sarkar, P.~Korshunov, L.~Colbois, and S.~Marcel.
\newblock Are {GAN}-based morphs threatening face recognition?
\newblock In {\em ICASSP 2022-2022 IEEE International Conference on Acoustics, Speech and Signal Processing (ICASSP)}, pages 2959--2963. IEEE, 2022.

\bibitem{scherhag2019detection}
U.~Scherhag, L.~Debiasi, C.~Rathgeb, C.~Busch, and A.~Uhl.
\newblock Detection of face morphing attacks based on {PRNU} analysis.
\newblock {\em IEEE Transactions on Biometrics, Behavior, and Identity Science}, 1(4):302--317, 2019.

\bibitem{scherhag2017biometric}
U.~Scherhag, A.~Nautsch, C.~Rathgeb, M.~Gomez-Barrero, R.~N. Veldhuis, L.~Spreeuwers, M.~Schils, D.~Maltoni, P.~Grother, S.~Marcel, et~al.
\newblock Biometric systems under morphing attacks: Assessment of morphing techniques and vulnerability reporting.
\newblock In {\em 2017 International Conference of the Biometrics Special Interest Group (BIOSIG)}, pages 1--7. IEEE, 2017.

\bibitem{scherhag2020deep}
U.~Scherhag, C.~Rathgeb, J.~Merkle, and C.~Busch.
\newblock Deep face representations for differential morphing attack detection.
\newblock {\em IEEE transactions on information forensics and security}, 15:3625--3639, 2020.

\bibitem{schroff2015facenet}
F.~Schroff, D.~Kalenichenko, and J.~Philbin.
\newblock Facenet: A unified embedding for face recognition and clustering.
\newblock In {\em Proceedings of the IEEE conference on computer vision and pattern recognition}, pages 815--823, 2015.

\bibitem{sohl2015deep}
J.~Sohl-Dickstein, E.~Weiss, N.~Maheswaranathan, and S.~Ganguli.
\newblock Deep unsupervised learning using nonequilibrium thermodynamics.
\newblock In {\em International conference on machine learning}, pages 2256--2265. PMLR, 2015.

\bibitem{song2020denoising}
J.~Song, C.~Meng, and S.~Ermon.
\newblock Denoising diffusion implicit models.
\newblock {\em arXiv preprint arXiv:2010.02502}, 2020.

\bibitem{szegedy2015going}
C.~Szegedy, W.~Liu, Y.~Jia, P.~Sermanet, S.~Reed, D.~Anguelov, D.~Erhan, V.~Vanhoucke, and A.~Rabinovich.
\newblock Going deeper with convolutions.
\newblock In {\em Proceedings of the IEEE conference on computer vision and pattern recognition}, pages 1--9, 2015.

\bibitem{Venkatesh2020FaceMA}
S.~Venkatesh, R.~Ramachandra, K.~Raja, and C.~Busch.
\newblock Face morphing attack generation and detection: A comprehensive survey.
\newblock {\em IEEE transactions on technology and society}, 2(3):128--145, 2021.

\bibitem{venkatesh2020can}
S.~Venkatesh, H.~Zhang, R.~Ramachandra, K.~Raja, N.~Damer, and C.~Busch.
\newblock Can {GAN} generated morphs threaten face recognition systems equally as landmark based morphs?-vulnerability and detection.
\newblock In {\em 2020 8th International Workshop on Biometrics and Forensics (IWBF)}, pages 1--6. IEEE, 2020.

\bibitem{wang2004image}
Z.~Wang, A.~C. Bovik, H.~R. Sheikh, and E.~P. Simoncelli.
\newblock Image quality assessment: from error visibility to structural similarity.
\newblock {\em IEEE transactions on image processing}, 13(4):600--612, 2004.

\bibitem{wolberg1998image}
G.~Wolberg.
\newblock Image morphing: a survey.
\newblock {\em The visual computer}, 14(8-9):360--372, 1998.

\bibitem{xiao2021tackling}
Z.~Xiao, K.~Kreis, and A.~Vahdat.
\newblock Tackling the generative learning trilemma with denoising diffusion {GANs}.
\newblock {\em arXiv preprint arXiv:2112.07804}, 2021.

\bibitem{yang2022auc}
T.~Yang and Y.~Ying.
\newblock Auc maximization in the era of big data and ai: A survey.
\newblock {\em ACM Computing Surveys}, 55(8):1--37, 2022.

\bibitem{zhang2024generalized}
H.~Zhang, R.~Ramachandra, K.~Raja, and C.~Busch.
\newblock Generalized single-image-based morphing attack detection using deep representations from vision transformer.
\newblock In {\em Proceedings of the IEEE/CVF Conference on Computer Vision and Pattern Recognition}, pages 1510--1518, 2024.

\bibitem{zhang2021mipgan}
H.~Zhang, S.~Venkatesh, R.~Ramachandra, K.~Raja, N.~Damer, and C.~Busch.
\newblock {MIPGAN}—generating strong and high quality morphing attacks using identity prior driven {GAN}.
\newblock {\em IEEE Transactions on Biometrics, Behavior, and Identity Science}, 3(3):365--383, 2021.

\bibitem{zhang2024diffmorpher}
K.~Zhang, Y.~Zhou, X.~Xu, B.~Dai, and X.~Pan.
\newblock Diffmorpher: Unleashing the capability of diffusion models for image morphing.
\newblock In {\em Proceedings of the IEEE/CVF Conference on Computer Vision and Pattern Recognition}, pages 7912--7921, 2024.

\bibitem{zhang2018unreasonable}
R.~Zhang, P.~Isola, A.~A. Efros, E.~Shechtman, and O.~Wang.
\newblock The unreasonable effectiveness of deep features as a perceptual metric.
\newblock In {\em Proceedings of the IEEE conference on computer vision and pattern recognition}, pages 586--595, 2018.

\bibitem{zope2017survey}
B.~Zope and S.~B. Zope.
\newblock A survey of morphing techniques.
\newblock {\em International Journal of Advanced Engineering, Management and Science}, 3(2):239773, 2017.

\end{thebibliography}
}

\end{document}